\documentstyle[12pt,epsf]{article}
\def\MAG{M_m}              \def\mag{\ifmmode\MAG\else$\MAG$\fi}
\def\MDEB{M_{\cal P}}      \def\mdeb{\ifmmode\MDEB\else$\MDEB$\fi}
\def\MEL{M_e}              \def\mel{\ifmmode\MEL\else$\MEL$\fi}
\def\PPBAR{\langle\bar\psi\psi\rangle}
                           \def\ppbar{\ifmmode\PPBAR\else$\PPBAR$\fi}
\def\SYM{{\cal D}_4^h}     \def\sym{\ifmmode\SYM\else$\SYM$\fi}
\def\LT{N_\tau}            \def\t{\ifmmode\LT\else$\LT$\fi}
\def\TC{T_c}               \def\tc{\ifmmode\TC\else$\TC$\fi}
\def\LZ{N_z}               \def\z{\ifmmode\LZ\else$\LZ$\fi}
         \def\etc{{\sl etc.\/}}
\def\etal{{\sl et al.\/}}     \def\ie{{\sl i.e.\/}}
\def\PR{{\sl Phys.\ Rev.\ }}  \def\PRL{{\sl Phys.\ Rev.\ Lett.\ }}
\def\PL{{\sl Phys.\ Lett.\ }} \def\NP{{\sl Nucl.\ Phys.\ }}
    \def\RMP{{\sl Rev.\ Mod.\ Phys.\ }}

\begin{document}
\thispagestyle{empty}
\mbox{} \hspace{10.0cm} September 1993\\
\mbox{} \hspace{10.6cm} HLRZ 61/93\\
\mbox{} \hspace{10.6cm} BI-TP 93/49\\
\mbox{} \hspace{10.6cm} FSU-SCRI-93-106\\

\begin{center}
\vspace*{1.0cm}
{{\large GLUEBALL-LIKE SCREENING MASSES\\
         IN PURE SU(3) AT FINITE TEMPERATURES}} \\
\vspace*{1.0cm}
{\large B. Grossman,$^1$
        Sourendu Gupta,$^{1,2}$
        U.M.~Heller$^3$
        and F.~Karsch$^{1,4}$\\
\vspace*{1.0cm}
{\normalsize
$\mbox{}^1$ {HLRZ, c/o KFA J\"ulich,
             D-52425 J\"ulich, Germany.}\\
$\mbox{}^2$ {TIFR, Homi Bhabha Road,
             Bombay 400005, India.}\\
$\mbox{}^3$ {SCRI, Florida State University, Tallahassee,
             FL 32306-4052, USA.}\\
$\mbox{}^4$ {Fakult\"at f\"ur Physik, Universit\"at Bielefeld,
             P.O.Box 100131, D-33501 Bielefeld, Germany;} \\
$\mbox{}^{}$ {and ITP, University of California, Santa Barbara,
             CA 93106-4030, USA.}}\\
\vspace*{2cm}
{\large \bf Abstract}}
\end{center}

\setlength{\baselineskip}{1.3\baselineskip}

We investigate the finite-temperature excitation spectrum in the gluon
sector of $SU(3)$ pure gauge theory through measurements of screening
masses in correlations of loop operators. We develop the classification
of such operators under the symmetry group of the `$z$-slice'. In the
confined phase of the theory, we find that the spectrum dynamically
realises the zero temperature symmetries. We observe a large thermal
shift of the $0^{++}$ glueball mass. In the deconfined phase, the
spectrum distinguishes between operators coupling to electrically
and magnetically polarised gluon fields. The former yields a screening
mass equal to the Wilson-line screening mass; the latter, a method
for the measurement of the magnetic mass in the high-temperature limit.

\newpage\setcounter{page}1

\section{Introduction}\label{sec:introduction}

The finite temperature deconfinement and chiral symmetry restoration
transition in QCD and pure $SU(3)$ gauge theory have been the subject of
continuing attention. Many detailed studies of the screening lengths
obtained from mesonic and baryonic screening correlation functions have
been performed recently \cite{hscreen,mtc,gup92b}.
It has become clear that there is a qualitative difference between the
hadronic excitation spectra above and below the phase transition temperature
\tc. For $T<\tc$, the excitation spectrum resembles that at zero
temperature--- there are (nearly) massless pions, and all the well-known
families of hadrons. For $T>\tc$, on the other hand, quarks are liberated.
As $T$ increases beyond \tc, the renormalised quark masses approach those
computed in resummed perturbation theory
\cite{qscreen}, and external currents with mesonic quantum numbers are
almost always correlated through the exchange of a weakly interacting
quark-antiquark pair (three weakly interacting quarks for baryons)
\cite{mtc}. The exception is in the channel with pion quantum numbers.
Here the inter-quark interaction remains quite strong even up to $2\tc$, but
various detailed measurements show the absence of a pion pole
\cite{gup92b}.

Nevertheless, there are strong spatial correlations between quarks
\cite{ber92w}. A recent computation \cite{koc92} relates these correlations
to a certain non-perturbative property of the gauge sector of the theory.
This is the observation that spatial Wilson loops exhibit area law behaviour
at all temperatures \cite{spw}. Non-perturbative physics has long been
observed for $T\to\tc^+$ in the pure gauge theory. This argument implies
that the only non-perturbative behaviour observed in the fermion sector of
QCD until now is related to these very same features. In an attempt to further
study the gauge sector of the theory, we have performed extensive studies of
certain screening correlation functions in the pure gauge theory. In the same
way that these are used to study the excitation spectrum in the fermionic
sector of QCD, we use them for a detailed study of the excitation spectrum in
the pure gauge theory, both above and below \tc.

Detailed knowledge of one aspect of the pure gauge spectrum is available.
In perturbation theory it is possible to compute the gluon propagator. From
such a computation of the polarisation tensor, the position of the pole for
the electric gluon propagator may be obtained. This is called the electric
mass of the gluon, \mel. The one-loop result has long been known \cite{gro81}.
For an $SU(N_c)$ gauge group with $N_f$ (massless) flavours of quarks,
\begin{equation}
\mel^2\;=\; c_e g^2T^2,\qquad\qquad c_e={1\over3}(N_c+N_f/2).
\label{eq:mel}\end{equation}
In this paper we are interested in the case $N_f=0$. It should be noted that,
at non-zero temperatures, the appearance of an electric or magnetic mass in
the gauge field propagator does not violate the usual Ward identities
\cite{gro81}.  Furthermore, it has been argued that the pole is a
gauge-invariant quantity although defined through a gauge dependent
correlation function \cite{rebhan}.

The electric gluon mass gives rise to the phenomenon of Debye screening of
colour charges. This is usually seen in the screening of the colour-singlet
potential between static quarks and the Debye screening mass can be shown to
be close to \mel{} in perturbation theory \cite{nadkarni}. It turns out to be
somewhat easier to measure the screening mass in Polyakov line correlations,
\mdeb. Again, in perturbation theory, it has been shown that
$\mdeb\approx2\mel$
\cite{nadkarni}. Close to \tc, \mdeb{} has been measured through zero-momentum
correlations of Polyakov loops \cite{ape}. Since this mass is defined through
a gauge invariant correlation function, it can be given a non-perturbative
meaning. It should be emphasised that screening masses of correlations between
gauge invariant objects can only be connected with electric and magnetic gluon
masses in perturbation theory. Thus screening masses are more general
observables, which in the high-temperature and weak-coupling limit, and only in
this limit, may be connected to the gluon electric and magnetic masses. This
point has been dealt with more extensively in \cite{nadkarni}.

Although the one-loop computation referred to earlier shows that the magnetic
part of the gauge field does not acquire a mass, it is possible to argue that
a two-loop evaluation reveals quite the opposite. At finite temperature,
therefore, the so-called magnetic mass, \mag, must be of the form
\begin{equation}
   \mag\;=\; c_m g^2 T,
\label{eq:mag}\end{equation}
where $c_m$ is a dimensionless constant. However, all higher loop order
computations contribute to the same order in \mag, showing that this quantity
cannot be perturbatively evaluated. Thus, the number $c_m$ above is a fully
non-perturbative quantity. In this argument we have neglected possible terms
in $\ln g$. That these terms can be resummed into a power of $g$ cannot be
disproved, and indeed, has been conjectured. There were attempts long ago to
measure $\mag/T$ in lattice simulations \cite{magmass}. We shall demonstrate
that one of the screening masses we observe, in the high-temperature limit,
goes to $2\mag$, and hence provides a measurement of \mag{} in the same way
that a measurement of \mdeb{} does for \mel. In this study we have not been
able to analyse the scaling behaviour of \mag.

Technically, our measurements are similar to zero-temperature determinations
of glueball masses. We construct spatial screening correlators between
colour-singlet pure gauge sources which are loops of different shapes. These
correlation functions are measured along the $z$-direction at finite
temperatures. The transverse slice does not carry representations of the
cubic group, as at $T=0$, but of the group \sym. We describe the
representation theory of \sym, as it applies to our measurements,
in section 2. The corresponding screening masses yield information on the
physical excitation spectrum of the high temperature theory in much the same
way as do the hadronic screening masses.

The organisation of this paper is the following. In the next section we
present the group theory relevant to our measurements. In section 3 we set out
details of our simulations and the measurement procedures. The results are
collected in the following section and conclusions presented in section 5.

\section{Symmetries of Screening Correlators}\label{sec:loopop}

In simulations of equilibrium field theories at  finite temperatures,
the rotational (Lorentz) symmetry of the $T=0$ theory is not realised.
This is reflected in practice by the fact that one uses lattices of size
$\t\times N^2\times\z$, where $\t<N\le\z$ and $\t=1/aT$ (here $a$ is the
lattice spacing). Correlations between operators
at different spatial separations measured on such an equilibrium system
reveal the spectrum of screening lengths in the theory. Thus, one measures
correlation functions of operators separated in the $z$-direction. As a
result, it is immaterial whether or not $N=\z$; and, in fact, it may be more
convenient to choose $\z>>N$. In these circumstances, it becomes necessary
to identify the symmetry group of the continuum and lattice theories on
the `$z$-slices', and perform the decomposition of operators into irreducible
representations of this group. This section summarises the results for
loop and link operators in a gauge theory.

\subsection{The Symmetry Group}

Since the correlations extend in the $z$ direction, the relevant symmetry is
of the `$z$-slice'. In the infinite volume, continuum limit, the group action
is that of the symmetries of a cylinder ${\cal C}=O(2)\times\{1,\sigma_z\}$.
$O(2)$ is a non-Abelian group with two one-dimensional representations denoted
$0_1$ and $0_2$ and two-dimensional representations labeled by the
angular momentum $l$ (taking integer values). The $z$ component of
the angular momentum can have the two values $\pm l$. The operator $\sigma_z$
denotes reflections $t\rightarrow -t$ about the $x$-$y$ plane,
and plays the role of a parity. We
denote its eigenvalues, $\{1,-1\}$, by the symbol $P$. The representations
of $O(3)$, describing the zero temperature glueballs, reduce under
${\cal C}$ as
\begin{eqnarray}
  l_{O(3)}^P\to 0_1^P+\sum_{i=1}^l i_{\cal C}^P
                  \qquad&{\rm for}&\quad P=(-1)^l \nonumber\\
  l_{O(3)}^P\to 0_2^P+\sum_{i=1}^l i_{\cal C}^P
                  \qquad&{\rm for}&\quad P=-(-1)^l.
        \label{eq:reduction}
\end{eqnarray}
On a finite lattice the symmetry of the $z$-slice is denoted by
$\sym={\cal D}_4\times\{1,\sigma_z\}$. ${\cal D}_4$ is the group of symmetries
of a square and has 8 elements in 5 conjugacy classes---
\begin{itemize}
\item[$E$:] {identity}
\item[$C_4$:] {2 rotations about the $t$-axis by $\pm \pi/2$}
\item[$C_4^2$:] {rotation about the $t$-axis by $\pi$}
\item[$\sigma$:] {2 reflections on the $x-t$ and $y-t$ planes}
\item[$\sigma^*$:] {2 reflections on the planes built from the $t$-axis
and the 2 diagonals in the $x-y$ plane.}
\end{itemize}

\noindent ${\cal D}_4$ has four one-dimensional irreducible representations
(irreps $R$) $A_1$, $A_2$, $B_1$ and $B_2$ and one two-dimensional irrep $E$.
The labelling of the irreps of the full symmetry group is as
$R^{PC}$ where $P$ stands for parity under $\sigma_z$ and $C$ for C-parity.
The lowest few representations of ${\cal C}$ reduce under \sym{} as
\begin{eqnarray}
    0_1^P \to A_1^P\qquad&\qquad& 0_2^P \to A_2^P \nonumber\\
    1^P \to E^P\qquad&\qquad& 2^P \to B_1^P + B_2^P\\
    3^P \to E^P\qquad&\qquad& 4^P \to A_1^P + A_2^P. \nonumber
        \label{eq:reduxion}
\end{eqnarray}
Thus, if the screening masses increase with increasing angular momentum, we
find inequivalent \sym{} representations only for $l\le2$. Higher angular
momenta replicate these representations.

In order to compare to lattice data at $T=0$, we also give the operators
that transform according to the symmetry group $O_h$ of a hypercubic
lattice in $x,y,t$-space. Rotational symmetry makes this the spatial
symmetry group \cite{ber83}. At any non-zero value of $T$, there is no
symmetry argument for the operators and masses to split according
to this classification. The reduction $O_h$ to \sym{} yields
\begin{eqnarray}
 [A_1^{+,C}]_{O_h}=[A_1^{+,C}]_{\sym}&,&\quad
 [A_1^{-,C}]_{O_h}=[A_2^{-,C}]_{\sym} \nonumber\\{}
 [A_2^{+,C}]_{O_h}=[B_1^{+,C}]_{\sym}&,&\quad
 [A_2^{-,C}]_{O_h}=[B_2^{-,C}]_{\sym}, \nonumber\\{}
 [E^{+,C}]_{O_h}=[A_1^{+,C}]_{\sym}+[B_1^{+,C}]_{\sym}&,&
 [E^{-,C}]_{O_h}=[A_2^{-,C}]_{\sym}+[B_2^{-,C}]_{\sym},\\{}
 [T_1^{+,C}]_{O_h}=[A_2^{+,C}]_{\sym}+[E^{-,C}]_{\sym}&,&
 [T_1^{-,C}]_{O_h}=[A_1^{-,C}]_{\sym}+[E^{+,C}]_{\sym}, \nonumber\\{}
 [T_2^{+,C}]_{O_h}=[B_2^{+,C}]_{\sym}+[E^{-,C}]_{\sym}&,&
 [T_2^{-,C}]_{O_h}=[B_1^{-,C}]_{\sym}+[E^{+,C}]_{\sym}. \nonumber
       \label{eq:redusion}
\end{eqnarray}

\subsection{Representations of Loop Operators}

\begin{table}[htb]
\begin{minipage}{0.49\textwidth}
\begin{center}\begin{footnotesize}\begin{math}
\begin{array}{|l|l||r|r|r|}    \hline
   O_h    &    $\sym$  & P^4_{xy} & P^4_{xt} & P^4_{yt}
                  \\   \hline \hline
 A_1^{++} &  A_1^{++}  &  1 & 1 &  1\\ \hline
 E^{++}   &  A_1^{++}  & -2 & 1 &  1\\ \cline{2-5}
          &  B_1^{++}  &  0 & 1 & -1\\ \hline\hline
 T_1^{+-} &  A_2^{+-}  &  1 & 0 &  0\\ \cline{2-5}
          &  E^{--}    &  0 & 1 &  0\\ \cline{3-5}
          &            &  0 & 0 &  1\\ \hline
\end{array}\end{math}\end{footnotesize}\end{center}
  \caption{The 4-link operators}
  \label{tab:4link}
\medskip\noindent
\end{minipage}
\begin{minipage}{0.49\textwidth}
\begin{center}\begin{footnotesize}\begin{math}
\begin{array}{|l|l||r|r|r|r|}    \hline
   O_h    &    $\sym$  & T^6_{xyt} & T^6_{xy-t} & T^6_{y-xt} & T^6_{y-x-t}
                  \\   \hline \hline
 A_1^{++} &  A_1^{++}  &  1 & 1 &  1&1   \\ \hline
 T_2^{++} &  B_2^{++}  &  1 & 1 &  -1& -1\\ \cline{2-6}
          &  E^{-+}    & 1  & -1& 0  &  0\\ \cline{2-6}
          &  E^{-+}    & 0  &  0& 1  & -1\\ \hline\hline
 A_2^{+-} &  B_1^{+-}  &  1 & 1 &  -1&-1   \\ \hline
 T_1^{+-} &  A_2^{+-}  &  1 & 1&1&1\\ \cline{2-6}
          &  E^{--}    & 1  &-1& 0 &  0\\ \cline{3-6}
          &            &  0 & 0 &  1& -1\\ \hline
\end{array}\end{math}\end{footnotesize}\end{center}
  \caption{The twisted 6-link operators}
  \label{tab:6linktwisted}
\medskip\noindent
\end{minipage}
\end{table}

\begin{table}[htb]
\begin{center}\begin{footnotesize}\begin{math}
\begin{array}{|l|l||r|r|r|r|r|r|}    \hline
   O_h  & $\sym$  & P^6_{xxy} & P^6_{xxt} & P^6_{yyx}& P^6_{yyt} & P^6_{ttx}
                & P^6_{tty}
                  \\   \hline \hline
 A_1^{++} &  A_1^{++}  &  1 & 1 &  1 & 1 & 1 & 1 \\ \hline
 E^{++}   &  A_1^{++}  & -2 & 1 & -2 & 1 & 1 & 1 \\ \cline{3-8}
          &            &  0 &-1 &  0 &-1 & 1 & 1 \\ \cline{2-8}
          &  B_1^{++}  &  0 & 1 &  0 &-1 & 1 &-1 \\ \cline{3-8}
          &            &  2 & 1 & -2 &-1 &-1 & 1 \\ \hline
 A_2^{++} &  B_1^{++}  & -1 & 1 &  1 &-1 &-1 & 1 \\ \hline\hline
 T_1^{+-} &  A_2^{+-}  &  1 & 0 & -1 & 0 & 0 & 0 \\ \cline{2-8}
          &  E^{--}    &  0 &-1 &  0 & 0 & 1 & 0 \\ \cline{3-8}
          &            &  0 & 0 &  0 & 1 & 0 &-1 \\ \hline
 T_2^{+-} &  B_2^{+-}  &  1 & 0 &  1 & 0 & 0 & 0 \\ \cline{2-8}
          &  E^{--}    &  0 &-1 &  0 & 0 &-1 & 0 \\ \cline{3-8}
          &            &  0 & 0 &  0 &-1 & 0 &-1 \\ \hline
\end{array}\end{math}\end{footnotesize}\end{center}
  \caption{The planar 6-link operators}
  \label{tab:6linkplanar}
\medskip\noindent
\end{table}

\begin{table}[htb]
\begin{center}\begin{footnotesize}\begin{math}
\begin{array}{|l|l||r|r|r|r|r|r|r|r|r|r|r|r|}    \hline
   O_h  & $\sym$  & B^6_{xyt}&B^6_{y-xt}&B^6_{-x-yt}&B^6_{-yxt}&
                  B^{6+}_{xty}   & B^{6-}_{xty}   &
                  B^{6+}_{yt-x}  & B^{6-}_{yt-x}  &
                  B^{6+}_{-xt-y} & B^{6-}_{-xt-y} &
                  B^{6+}_{-ytx}  & B^{6-}_{-ytx}
                  \\   \hline \hline
 A_1^{++} &  A_1^{++}
          &  1& 1 & 1 & 1 & 1 &0  &1  &0  &1  & 0  &1  & 0 \\ \hline
 E^{++}   &  A_1^{++}
          &  -4& -4 &-4  &-4  &2  &0  &2  &0  &2  & 0  &2  &0  \\
             \cline{2-14}
          &  B_1^{++}
          &  0& 0& 0&0 & 1 & 0 &-1& 0& 1 & 0 & -1 & 0   \\ \hline
 T_2^{++} &  B_2^{++}
          &  1&-1&1&-1& 0&0&0&0&0&0&0&0       \\ \cline{2-14}
          &  E^{-+}
          &  0&0&0&0&0&  1&0&0&0& -1& 0& 0      \\ \cline{3-14}
          &
          &  0&0&0&0&0&0&0& 1&0&0&0& -1        \\ \hline
 T_2^{-+} &  B_1^{-+}
          &  0&0&0&0&0&  1& 0& -1&0& 1& 0& -1       \\ \cline{2-14}
          &  E^{++}
          &  1&-1&-1&1& 0& 0& 1 &0&0&0&-1&0      \\ \cline{3-14}
          &
          &  1 &1 & -1&-1&-1& 0&0&0& 1& 0&0&0        \\ \hline
 T_1^{-+} &  A_1^{-+}
          &  0&0&0&0&0&  1& 0& 1&0& 1& 0& 1       \\ \cline{2-14}
          &  E^{++}
          &  -1&1&1&-1 &0& 0& 1 &0&0&0&-1&0      \\ \cline{3-14}
          &
          &  1 &1 & -1&-1& 1& 0&0&0& -1& 0&0&0        \\ \hline\hline
 A_1^{--} &  A_2^{--}
          &  1& 1 & 1 & 1 & 0& 1 &0  &1  &0  &1  & 0  &1 \\ \hline
 E^{--}   &  A_2^{--}
          &  2&2&2&2& 0&1&0&1&0&1&0&1 \\
             \cline{2-14}
          &  B_2^{--}
          &  0& 0& 0&0&0  & 1 & 0 &-1& 0& 1 & 0 & -1   \\ \hline
 T_2^{--} &  B_1^{--}
          &  1&-1&1&-1& 0&0&0&0&0&0&0&0       \\ \cline{2-14}
          &  E^{+-}
          &  0&0&0&0&  1&0&0&0& -1& 0& 0 &0     \\ \cline{3-14}
          &
          &  0&0&0&0&0&0& 1&0&0&0& -1 &0       \\ \hline
 T_2^{+-} &  B_2^{+-}
          &  0&0&0&0&  1& 0& -1&0& 1& 0& -1 &0      \\ \cline{2-14}
          &  E^{--}
          &  -1&1&1&-1&0&0&0&1&0&0&0&-1      \\ \cline{3-14}
          &
          &  -1&-1&1&1&0&-1&0&0&0&1&0&0       \\ \hline
 T_1^{+-} &  A_2^{+-}
          &  0&0&0&0&  1& 0& 1&0& 1& 0& 1 &0      \\ \cline{2-14}
          &  E^{--}
          &  1&-1&-1&1& 0&0& 0& 1 &0&0&0&-1      \\ \cline{3-14}
          &
          &  -1 &-1 & 1&1& 0&1&0&0&0& -1& 0&0        \\ \hline
\end{array}\end{math}\end{footnotesize}\end{center}
  \caption{The bent 6-link operators}
  \label{tab:6linkbent}
\medskip\noindent
\end{table}

\begin{table}[htb]
\begin{center}\begin{footnotesize}\begin{math}
\begin{array}{|l||r|r|r|r|r|r|r|r|}    \hline
    $\sym$   & T^{8P}_{xy-xt} & T^{8P}_{y-x-yt} & T^{8P}_{-x-yxt}
           & T^{8P}_{-yxyt} & T^{8P}_{x-y-xt} & T^{8P}_{yx-yt}
           & T^{8P}_{-xyxt} & T^{8P}_{-y-xyt}
                  \\   \hline \hline
 A_1^{P} & 1&1&1&1&1&1&1&1  \\ \hline
 B_1^{P} & 1&-1&1&-1&1&-1&1&-1  \\ \hline
 A_2^{P} & 1&1&1&1&-1&-1&-1&-1  \\ \hline
 B_2^{P} & 1&-1&1&-1&-1&1&-1&1  \\ \hline
 E^{P}   & 1&0&-1&0&0&1&0&-1  \\ \cline{2-9}
         & 0&1&0&-1&-1&0&1&0  \\ \cline{2-9}
         & 1&0&-1&0&0&-1&0&1  \\ \cline{2-9}
         & 0&1&0&-1&1&0&-1&0  \\ \hline
\end{array}\end{math}\end{footnotesize}
\end{center}
  \caption{The combinations of 8-link operators which yield the representations
           shown; $P$ takes the values $\pm1$.}
  \label{tab:8link}
\medskip\noindent
\end{table}

In this subsection we list the loop operators required to build the
different representations of \sym. We give all the 4- and 6-link operators,
as well as one group of 8-link operators which is required to build the
representation $A_2^{-+}$. This representation cannot be built out of the
shorter loops. The complete list of loops used is
\begin{eqnarray}
  P^4_{\mu\nu}&\;=\;{1\over2N} {\rm tr}\, U(\mu,\nu,-\mu,-\nu)
   & \mbox{ 4-link} \nonumber\\
  T^6_{\mu\nu\rho}&\;=\;{1\over2N} {\rm tr}\,U(\mu,\nu,\rho,-\mu,-\nu,-\rho)
   & \mbox{ twisted 6-link} \nonumber\\
  P^6_{\mu\mu\nu}&\;=\;{1\over2N} {\rm tr}\,U(\mu,\mu,\nu,-\mu,-\mu,-\nu)
   & \mbox{ planar 6-link}\\
  B^6_{\mu\nu\rho}&\;=\;{1\over2N} {\rm tr}\,U(\mu,\nu,\rho,-\nu,-\mu,-\rho)
   & \mbox{ bent 6-link} \nonumber\\
  T^8_{\mu\nu-\mu\rho}&\;=\;{1\over2N}
    {\rm tr}\,U(\mu,\nu,-\mu,-\nu,-\mu,\rho,\mu,-\rho)
   & \mbox{ twisted plaquette pair}. \nonumber
\label{eq:list}\end{eqnarray}
The notation $U(\mu,\nu,\cdots)$ denotes a product of link matrices $U$ over
a loop. The loop is specified from an arbitrary starting point by the
directions $\mu$, $\nu$, \etc, in order. The further definitions
\begin{eqnarray}
  B^{6\pm}_{\mu,t,\nu}&=&B^6_{\mu,t,\nu}\pm B^6_{\mu,-t,\nu} ~, \nonumber\\
  T^{8\pm}_{\mu,\nu,-\mu,t}&=&T^8_{\mu,\nu,-\mu,t}\pm T^8_{\mu,\nu,-\mu,-t}
     \label{eq:newop}
\end{eqnarray}
turn out to be convenient. The $C=1$ representations are always obtained
by taking the real part of the loop and the $C=-1$ representations by the
imaginary part.

The tables can be used to get two related pieces of information. The entries
are the coefficients for linear combinations of loop operators required to
obtain the indicated representation of \sym. The combinations for 4-link
operators are given in Table~\ref{tab:4link}; for twisted 6-link operators
in Table~\ref{tab:6linktwisted}; for planar and bent 6-link operators in
Tables~\ref{tab:6linkplanar} and \ref{tab:6linkbent} respectively; and
finally for a set of 8-link operators in Table~\ref{tab:8link}. For 4- and
6-link operators, the breakup of $O_h$ representations (the $T=0$ symmetry
group) for these sets of operators into representations of \sym{} are also
specified. For the 8-link operators this decomposition is not given. In our
later analysis we shall require this information for the $B_1^{++}$ and
$B_2^{++}$ representations built out of these operators. In these two cases,
the containing representations of $O_h$ are uniquely fixed to be $E^{++}$ and
$T_2^{++}$.

\subsection{Link Operators and their Direct Products}

\begin{table}[htb]\begin{center}\begin{footnotesize}\begin{math}
\begin{array}{|l||l|l|}    \hline
D_1 \otimes D_2  &  D &  \mbox{state vector}   \\\hline\hline
A_1^{--} \otimes A_1^{--} & A_1^{++}          & A_t A_t      \\\hline
A_1^{--} \otimes E^{+-}   & E^{-+}            & A_t A_x + A_x A_t \\
                          &                   & A_t A_y + A_y A_t \\\hline

E^{+-} \otimes E^{+-}     & A_1^{++}          &A_x^2+A_y^2  \\
                          & \oplus B_1^{++}   & A_x^2 - A_y^2      \\
                          & \oplus B_2^{++}   & A_x A_y  +A_y A_x   \\
                          & (\oplus A_2^{++}) & (A_x A_y -A_y A_x)   \\\hline
\end{array}\end{math}\end{footnotesize}\end{center}
\caption{
  Irreducible components $D$ of twofold products $D_1\otimes D_2$ of
  representations of \sym. Note that the $A_2^{++}$ state vector cannot
  be combined into a colourless symmetric state and is therefore forbidden.}
\label{tab:2gsym}\medskip\noindent\end{table}

At high temperatures one may expect the observed correlations to be explained
in terms of multiple gluon exchanges. Therefore it is necessary to perform a
reduction of multigluon states to the irreps of \sym. In order to derive the
corresponding selection rules we start from the definition of a gluon field on
the lattice---
\begin{equation}
A_\mu(x)={1\over2i}\bigg[\big\{U_\mu(x)-U_\mu^\dagger(x)\big\}-
     {\rm tr}\big\{U_\mu(x)-U_\mu^\dagger(x)\big\}\bigg]
\label{eq:glueop}\end{equation}
for a gluon operator in any appropriately fixed gauge. Thus the electric gluon,
$A_t$, transforms according to the irrep $A_1^{--}$ of \sym{} and the magnetic
gluons, $A_x$ and $A_y$, according to the irrep $E^{+-}$ of \sym.

The reduction of the direct product representations is  straightforward.
For two-gluon states the results are listed in Table \ref{tab:2gsym}. Since
each gluon is a colour octet, there is only one symmetric combination of
two gluons to a colour singlet. Hence the $A_2^{++}$ state of
Table~\ref{tab:2gsym} cannot be combined into a colourless state. However,
the $A_1^{++}$, $E^{-+}$, $B_1^{++}$, and $B_2^{++}$ are allowed as two
gluon states. The reduction of the three gluon states is listed in
Table~\ref{tab:3gsym}.

\begin{table}[htb]\begin{center}\begin{footnotesize}\begin{math}
\begin{array}{|l||l|}    \hline
D_1 \otimes D_2 \otimes D_3 &  D   \\ \hline\hline
A_1^{--} \otimes A_1^{--} \otimes A_1^{--}  &    A_1^{--}  \\ \hline
A_1^{--} \otimes A_1^{--} \otimes E^{+-}    &    E^{+-}   \\  \hline
A_1^{--} \otimes E^{+-}   \otimes E^{+-}    &    A_1^{--}+A_2^{--}+B_1^{--}
                                                     +B_2^{--} \\ \hline
E^{+-}   \otimes E^{+-}   \otimes E^{+-}    &   4 E^{+-}  \\ \hline
\end{array}\end{math}\end{footnotesize}\end{center}
\caption{
 Irreducible components $D$ of threefold products $D_1\otimes D_2\otimes D_3$
 of representations of $\SYM$.}
\label{tab:3gsym}\medskip\noindent\end{table}

We note that the reduction of two and three gluon states in Tables
\ref{tab:2gsym} and \ref{tab:3gsym} does not take into account the possible
relative motion of the gluons. It thus applies directly only to gluons that
are in an orbital s-state. We were able to observe an $A_2^{-+}$ state (see
section 4), which does not appear in Tables \ref{tab:2gsym} and
\ref{tab:3gsym}. It is easy to see that it can be built of two gluons in an
orbital p-state.

The two possible channels in the $A_1^{++}$ sector can thus be separated
by splitting operators into two groups--- those containing links in the
time direction, and those not. These two sets should give, in general,
unequal screening masses. In the $g(T)\to0$ limit, these screening
masses are then equal to 2\mel{} and 2\mag. In this limit the $B_1^{++}$
and $B_2^{++}$ screening masses should also yield 2\mag; although for
any finite $T$ there should be splittings between these three screening
masses. Similarly, the $A_2^{-+}$ masses should be asymptotically
degenerate with these masses. In the $T\to\infty$ limit, obvious cross
checks should be provided by the screening masses in the $E^{-+}$ and
the negative C-parity channels.

\section{Simulations and Measurements}\label{sec:simulations}

Simulations were carried out in pure $SU(3)$ gauge theory at two temperatures.
One was chosen to be in the confined phase, at about $0.75\tc$, and the other
in the deconfined phase, at about $1.5\tc$. Since the phase transition for
$\t=6$ occurs at $\beta_c\simeq5.9$ \cite{iwa92:histnp}, we chose to
work with $8\times16^3$ and $4\times16^3$ lattices at fixed $\beta=5.93$.
Scaling was investigated by a separate run on a $6\times16^3$ lattice at
$\beta=6.101$. Both the $\t=4$ and 6 lattices were at the same physical
temperature in the deconfined phase. Further details of the runs are listed
in Table~\ref{tab:statist}.

\begin{table}[htb]
\begin{center}\begin{footnotesize}\begin{math}
\begin{array}{|l||r|r|r|}    \hline
  N_{\tau}\times N_{\sigma}^3 &  \beta & N_m &   N_b  \\ \hline
  4\times 16^3    &  5.93 & 13000 &  1300  \\ \hline
  8\times 16^3    &  5.93 &  5000 &   500  \\ \hline
  6\times 16^3    &  6.101&  6300 &   630  \\ \hline
\end{array}\end{math}\end{footnotesize}\end{center}
\caption{
  Lattice sizes $N_{\tau}\times N_{\sigma}^3$, Wilson coupling $\beta$, number
  of measurements $ N_m $ and size $N_b$ of one jack-knife bin for analysis.}
\label{tab:statist}\medskip\noindent\end{table}

The simulations were performed with an $SU(2)$ subgroup over-relaxation (OR)
mixed with a Cabbibo-Marinari pseudo-heatbath (HB) update \cite{cab82:hist}.
We used 4 steps of OR to each of HB. The HB step used a Kennedy-Pendleton
$SU(2)$ heatbath \cite{ken85:histnp}. The time for 4 OR + 1 HB updates is 115
$\mu$s per link on a Cray Y-MP. Measurements were performed on configurations
separated by 50 HB steps. The auto-correlations times away from \tc{} are
expected to be rather small, and each of these configurations is essentially
uncorrelated with the others.

\subsection{Fuzzed Operators}

Past experience, at $T=0$, indicates that the signal to noise ratio for
loop-loop correlations decays extremely fast with separation. It seems
necessary to use a signal enhancement technique in order to measure
correlation functions at distances greater than 2 or 3 lattice spacings.
Accordingly, we used loops constructed out of fuzzed links \cite{tep86}.
This is a well-known technique. For each configuration, the link matrices
generated are considered to be the level $l=0$ of fuzzing. Then a link
$U^l_\mu(i)$ at fuzzing level $l$ is given in terms of those at the previous
level, $l-1$, by the relation
\begin{eqnarray}
   U^l_\mu(i) \;=\; \left[ U^{l-1}_\mu(i)\right.&& U^{l-1}_\mu(i+\mu)
       \nonumber\\
      +&&\sum_{\nu \ne \mu}  \left. U^{l-1}_\nu(i) U^{l-1}_\mu(i+\nu)
                   U^{l-1}_\mu(i+\nu+\mu)
            \left(U^{l-1}_\nu(i+2\mu)\right)^\dagger  \right]_{SU(3)},
\label{eq:fuzz}\end{eqnarray}
where the subscript denotes a projection onto $SU(3)$. This projection
is obtained through the polar decomposition of the general complex
$3\times3$ matrix, $M$, on the right--- $M=\omega U H$. $H$ is a
Hermitian matrix, $\omega$ a complex number and $U$ the required special
unitary matrix. We let the indices $\mu$ and $\nu$ in eq.\
(\ref{eq:fuzz}) range over the $x$, $y$ and $t$ directions. In each
direction we perform fuzzing up to level $(\log_2 N-1)$, where $N$ is the
lattice size. Thus, for $\t=4$ we perform fuzzing up to level 1 for
links in the $t$ direction.

We constructed the operators listed in section 2 out of the fuzzed links
obtained by this procedure. Note also that the exclusion of the $z$-direction
links from the sum in eq.\ (\ref{eq:fuzz}) implies that the separation in the
correlation functions measured is a well-defined quantity. Further, since the
$z$ links do not enter any of the operators we study, the loss in fuzzing is
probably not very important.

\subsection{Mass Measurements}

Since different operators realising the same representation of \sym{} can be
correlated through the exchange of the same state, we have measured all
elements of the full correlation matrix
\begin{equation}
  C_{\alpha,\beta}(z) \;=\; \langle L_\alpha(0)L_\beta(z)\rangle -
                         \langle L_\alpha\rangle\langle L_\beta\rangle,
\label{eq:full}\end{equation}
where $\alpha$ and $\beta$ label the loop operators in a given symmetry
sector. The loop operators were summed over all sites on a $z$-slice, in
order to project onto zero momentum. We can obtain the lowest lying
excitation by a variational procedure \cite{lue90}. This involves solving
the generalised eigenvalue problem
\begin{equation}
C_{\alpha,\beta}(z_1)Y_\beta=\lambda(z_1,z_0) C_{\alpha,\beta}(z_0) Y_\beta.
\label{eq:var}\end{equation}
Then the quantity
\begin{equation}
  g(z)\;=\; Y_\alpha C_{\alpha,\beta}(z) Y_\beta,
\label{eq:projcf}\end{equation}
where $Y$ belongs to the lowest eigenvalue, projects onto the correlation
function with the smallest mass. Local masses, $m(z)$, were extracted by
solving the equation
\begin{equation}
{g(z)\over g(z-1)}\;=\;
  {\cosh\left[m(z)(\z/2-z)\right]\over\cosh\left[m(z)(\z/2-(z-1))\right]},
\label{eq:localmass}\end{equation}
where the left hand side was obtained from the measurements. Jack-knife
estimators were used for the local masses and their errors. If the
diagonalisation procedure exactly isolates the lowest mass into $g(z)$,
then $m(z)$ should be independent of $z$ within errors. In practise this
does not happen; and we have to identify the lowest mass through a plateau
in $m(z)$. The effectiveness of the diagonalisation procedure is nevertheless
seen in the extension of this plateau region.

We restricted ourselves to diagonalising for $z_1=1$ and 2 at $z_0=0$ in order
to determine the leading eigenvector $Y$ for each set of quantum numbers. The
two different sets of eigenvectors were found to be reasonably similar. It
is useful to recall that this eigenvector gives the weight of each loop to
the lowest state, and hence is an analogue of the modulus squared of the
``wavefunction''.

\subsection{Polyakov Loops}

We have also measured the zero-momentum correlation functions between
Polyakov loops
\begin{equation}
  C_{ww}(z)\;=\;\langle\Omega^\dagger(z)\Omega(0)\rangle -
                \langle\Omega^\dagger\rangle\langle\Omega\rangle.
\label{eq:ldagl}\end{equation}
The zero-momentum projection is achieved through the use of the walls
\begin{equation}
\Omega(z)\;=\;\sum_{xy}\Omega(x,y,z),\qquad
        \Omega(x,y,z)\;=\;{\rm tr}\,\prod_t U_0(x,y,z,t).
\label{eq:polyakov}\end{equation}
The conjugate wall operator $\Omega^\dagger$ is defined by replacing
the link operators in eq.\ (\ref{eq:polyakov}) by their Hermitian
conjugates. The static potential $V(\!\vec{\,r})$ between a quark and an
antiquark at distance $\vec{r}=(x,y,z)$ is usually measured through
point to point correlation functions in the form
\begin{equation}
\exp (-V(\!\vec{\,r})/T)\;=\;
   {\langle\Omega^\dagger(\!\vec{\,r})\Omega(0)\rangle\over
    \langle\Omega^\dagger\rangle\langle\Omega\rangle}.
\label{eq:potential}\end{equation}
In perturbation theory the colour averaged potential above is expected
to have the form $\exp(-\mdeb r)/r^2$. Expanding the exponential
in eq.\ (\ref{eq:potential}) and substituting the perturbative form for
the potential, one can then derive the appropriate form for $C_{ww}$. In
the long distance limit it is given by the exponential integral
$E_1(\mdeb z)$. We actually use the symmetrised form
\begin{equation}
C_{ww}(z) \propto E_1(\mdeb z)+E_1(\mdeb(\LZ- z)),
\label{eq:corrscreensym}\end{equation}
and determine \mdeb{} by an appropriate generalisation of a local mass.

In the confining phase, $V(\!\vec{\,r})\propto rK_{eff}/T$. The appropriate
symmetrised form for the zero momentum correlation function is then
\begin{equation}
C_{ww}(z)\;=\;
    A\cosh\left[{K_{eff}\over T}\left({\z\over2}-z\right)\right],
\label{eq:corrconf}\end{equation}
where $A$ is some constant. It is clear that a local mass measurement
yields $K_{eff}/T$ and hence a plateau in this can be used for a
measurement of the string tension at finite temperatures.

The correlation function
\begin{equation}
 C_{aa}(z)=\langle|\Omega^\dagger(z)||\Omega(0)|\rangle -
           \langle|\Omega^\dagger|\rangle\langle|\Omega|\rangle,
 \label{eq:polabs}
\end{equation}
on the other hand, has trivial triality, and hence furnishes an alternative
measurement of the mass in the $A_1^{++}$ channel.

\section{Results}

The local masses obtained from eqs.~(\ref{eq:var}) through~(\ref{eq:localmass})
are listed in Table~\ref{tab:localz}. We have underlined the values which
we consider to be the best estimates of the asymptotic masses. In all our
measurements an $A_1^{++}$ mass turned out to be the lowest. The
corresponding correlation functions were the least noisy, and the local
masses could be followed out to distance 5 or 6. In this case it seems
quite likely that we are able to identify the lowest mass in this channel.
Apart from this, local masses at distance greater than 1 could only be
measured with reasonable accuracy in the channels with quantum numbers
$A_2^{-+}$, $B_1^{++}$ and $B_2^{++}$. In these cases we could follow the
local masses up to about distance 3. It is likely that the extraction of a
mass in these latter channels is contaminated by the presence of higher
states. There is more structure in our measurements than is seen in the
group theory presented in section 2. This is good news, since it indicates
non-trivial dynamics.

\begin{table}[htb]
\begin{center}\leavevmode\begin{footnotesize}\begin{math}
\begin{array}{|l|l||l|l|l||l|l|l|l|l|l|}    \hline
N_\tau&\beta&\mbox{rep}&\mbox{operator}&l&\multicolumn6{|c|}z\\
 \cline{6-11}&&&&&1&2&3&4&5&6\\\hline\hline
8&5.93 &A_1^{++}&\mbox{local, }A_1^{++}&2&0.87(1)&0.78(2)& 0.77(4)&0.7(1)
                &\underline{0.46(8)}&0.5(2)\\\cline{3-11}
      &&A_1^{++}&\mbox{local, } E^{++} &2&1.39(2)&1.07(5)&\underline{1.0(2)}
                &0.6(3)&&\\ \cline{3-11}
      &&A_1^{++}&C_{aa}&2&1.23(2)&0.63(3)&0.58(5)&\underline{0.51(8)}
                &0.6(1)&0.6(2)\\ \cline{3-11}
              &      & A_2^{-+} & \mbox{local} & 2 &
                1.84(2)&1.6(1)&\underline{1.2(5)}&&&\\ \cline{3-11}
              &      & B_1^{++} & \mbox{local, } E^{++} & 2 &
                1.38(2)&1.21(6)&\underline{1.0(1)}&0.8(3)&&\\ \cline{3-11}
              &      & B_2^{++} & \mbox{local, } T_2^{++} & 2 &
                1.45(2)&1.27(6)&\underline{1.2(1)}&1.2(4)&&\\ \hline
4 & 5.93 & A_1^{++} & \mbox{local, t} & 2 &
                1.012(6)&0.80(3)&\underline{0.73(5)}&0.8(1)&0.7(3)&
                \\ \cline{3-11}
              &      & A_1^{++} & \mbox{local, no t} & 2 &
                1.36(1)&1.27(4)&\underline{1.3(1)}&0.7(2)&&
                \\ \cline{3-11}
              &      & A_2^{-+} & \mbox{local} & 2 &
                1.54(1)&1.12(3)&\underline{1.0(1)}&0.8(2)&&\\ \cline{3-11}
              &      & B_1^{++} & \mbox{local} & 2 &
                2.22(2)&\underline{2.0(2)}&&&&\\ \cline{3-11}
              &      & B_2^{++} & \mbox{local} & 2 &
                2.37(2)&\underline{1.9(2)}&&&&\\ \hline
6 & 6.101& A_1^{++} & \mbox{local, t} & 2 &
                0.97(1)&0.64(4)&0.61(8)&0.51(13)&\underline{0.43(11)}&
                \\ \cline{3-11}
              &      & A_1^{++} & \mbox{local, no t} & 2 &
                1.063(7)&0.96(2)&\underline{0.97(7)}&1.1(2)&&
                \\ \cline{3-11}
              &      & A_1^{++} &C_{aa}& 0 &
                1.0(1)&0.56(1)&0.46(2)&0.51(4)&\underline{0.49(7)}&0.48(6)
                \\ \cline{3-11}
              &      & A_2^{-+} & \mbox{local} & 2 &
                1.40(2)&1.00(3)&\underline{0.83(6)}&0.8(2)&&\\ \cline{3-11}
              &      & B_1^{++} & \mbox{local} & 2 &
                1.82(1)&1.6(1)&\underline{1.3(2)}&&&\\ \cline{3-11}
              &      & B_2^{++} & \mbox{local} & 2 &
                1.82(2)&1.57(7)&\underline{1.5(5)}&&&\\ \hline
\end{array}\end{math}\end{footnotesize}
\caption{
   Local masses, $m(z)$, for the different representations tabulated as a
   function of $z$. We take the underlined values as our best estimate for
   the asymptotic masses. In the confined phase ($\t=8$) we list the $O_h$
   representation to which the operator belongs, while in the deconfined
   phase ($\t=4$ and 6) we indicate for the representation whether the
   operator contains time-like links or not.}
\label{tab:localz}\end{center}\end{table}

\subsection{The $A_1^{++}$ Channel}

Note that the $A_1^{++}$ mass obtained through the trivial-triality
correlations of Polyakov loops is always equal to the lowest mass in
the $A_1^{++}$ channel extracted using loop operators. This provides a
cross check on the proper extraction of this mass.

Below \tc{} this channel descends from the $E^{++}$ and $A_1^{++}$
representations of the $O_h$ group. Grouping the operators constituting
this representation of \sym{} into two sets, one for each of the $T=0$
symmetries, we find that the lowest mass in the former group is about
twice that in the latter. Such a splitting is also indicated by the
eigenvectors, $Y_\beta$, defined in eq.\ (\ref{eq:var}). The
low-temperature system thus seems to realise the $T=0$ symmetries
dynamically.

This is most easily demonstrated by the eigenvectors restricted to the
two dimensional space spanned by plaquette correlations, where the
plaquette operators could have a link in the Euclidean-time direction
($P_t=P^4_{xt}+P^4_{yt}$) or no links in this direction ($P_s=P^4_{xy}$).
An operator in the
$E^{++}$ representation of $O_h$ can be formed by the combination $-2P_s
+P_t$. We denote this in the shorthand notation ($-2$,1). The $A_1^{++}$
representation of $O_h$ is given by (1,1). Diagonalisation of the
correlation matrix in this subspace gives the lowest mass of $0.5\pm0.1$
with the corresponding eigenvector equal to ($0.91\pm0.03$,$1.00\pm
0.03$). The next mass is roughly 1.0 and the eigenvector corresponding
to this is ($-2.093\pm0.007$,$1.00\pm0.02$). Similar results are
obtained in the three dimensional subspace of planar 6-link operators.

Since the $T=0$ symmetries seem to be generated dynamically, it is of
interest to ask about the temperature dependence of glueball masses
below \tc. Our best estimate for
the $A_1^{++}$ ($O_h$ representation) mass at $T=0.75T_c$ is 0.49(8).
This mass has been measured at $T=0$
\cite{mic89} to be $0.81(3)$ at $\beta=5.93$ on a $24^3\times36$
lattice. Measurements at $\beta=5.90$ on $12^4$ lattices verify that
finite size effects near this coupling are small. Our measurement
yields
\begin{equation}
{m(T=0.75T_c)\over m(T=0)}\;=\; 0.6\pm0.1
              \qquad\qquad(A_1^{++},\;\beta=5.93).
\label{eq:ratio}\end{equation}
Thus, the thermal shift is much larger than that seen in quenched hadron
masses \cite{hadron}.

Above \tc, the situation is quite different. There is no evidence of a
splitting of the $A_1^{++}$ level according to the $T=0$ symmetry. On the
other hand, a definite splitting is observed when the operators are grouped
into two sets, one with loops containing no links in the $t$ direction, and
the other containing at least one such link. From the decomposition of the
multi-gluon states (see Table\ \ref{tab:2gsym}) it is seen that these two
sets correspond to the exchange of two magnetic and two electric gluons
respectively. This interpretation is further supported by the agreement of
the second of these two masses with \mdeb. It should be noted, however,
that loops containing both space-like and time-like links can couple to
magnetic and electric gluons with the lighter ones dominating the
correlation function at large distances. For our parameters the electric
gluons seem to be lighter and hence dominating.

The eigenvectors also give evidence of such a decomposition. The two
dimensional space of plaquette correlations for the $\t=6$ lattice again
demonstrates this very well. The lowest mass corresponds to the
eigenvector ($0.2\pm0.2$,$1.01\pm0.05$), \ie, to an operator with large
overlap with $P_t$. The next eigenvector seems to have a larger overlap
with $P_s$ than with $P_t$. Similar results can also be obtained in the
three dimensional space of planar 6-link operators. Finally, note that
$m/T$ is the same for the $\t=4$ and 6 lattices, showing that finite
lattice-spacing effects are under control.

\subsection{Polyakov Loops}

The results for the effective string tension $K_{eff}$ below \tc\ and the
 mass $\mdeb$ above \tc\ are listed in Table~\ref{tab:stringfit}. For the
effective string tension at a temperature $T=0.75\tc$ we find
$\sqrt{K_{eff}(T)}=0.197(6).$
Although there is no measurement of $\sqrt{K_{eff}}$ at precisely this
coupling at $T=0$, it is possible to estimate this value from an
interpolation of the data given in \cite{mic89}. We find
\begin{equation}
 \sqrt{K_{eff}(T=0.75\tc)\over K_{eff}(T=0)}\;=\;0.83(4).
\label{eq:tdepkeff}\end{equation}
Consequent to this we find that the ratio $m(A_1^{++})/\sqrt{K_{eff}}$
is itself temperature dependent, changing from $3.5(2)$ at $T=0$ to a
value of $2.5(5)$ at $T=0.75\tc$.

\begin{table}[htb]\begin{center}\leavevmode\begin{footnotesize}\begin{math}
\begin{array}{|l|l||l||l|l|l|l|l|l|l|}    \hline
N_t&\beta&l&\multicolumn7{|c|}z\\\cline{4-10}&&&1&2&3&4&5&6&7\\\hline\hline
8&5.93 &2&0.071(1)&0.043(1)&0.041(1)&0.040(3)&\underline{0.039(3)}
             &0.036(4)&0.043(3)\\\hline
4&5.93 &0&&0.400(8)&0.47(3)&\underline{0.51(5)}&0.66(12)&0.5(2)&\\\hline
6&6.101&0&&0.241(8)&0.24(1)&0.34(4)&0.34(6)&\underline{0.33(6)}&0.3(1)\\\hline
\end{array}\end{math}\end{footnotesize}\end{center}
\caption{Local values of $K_{eff}$ for $\t=8$ and $\mdeb$ for $\t=4$ and 6.}
\label{tab:stringfit}\end{table}

The simulations for $\t=4$ and 6,
corresponding to a temperature of about $1.5\,\tc$, yield consistent values for
$\mdeb/T$. From the data in the table, it is seen that we obtain
\begin{equation}
\mdeb=2.0(2)T \,\mbox{ at }\,\LT=4\mbox{ and }\qquad
\mdeb=2.0(4)T \,\mbox{ at } \LT=6.
\label{eq:debeyeres}\end{equation}
These should be compared to the value $1.63(8)$, obtained at $2\tc$ in
\cite{debye} from measurements of the static
inter-quark potential, and the one-loop perturbation theory result
$2\,\mel/T=2\,g(T)$ where $g(T)\approx 1$.

\subsection{Other Channels}

At $T=0$ the measurement of excited glueball masses is not easy. Nor
does a finite temperature render this measurement any easier.
Correlation functions in channels other than $A_1^{++}$ cannot be
followed very far. As a result, evidence for a genuine plateau in the
local masses is not conclusive. Although we are able to tentatively
assign screening masses to correlations in these channels, more detailed
work will be necessary to achieve the reliability obtained in the
$A_1^{++}$ channel.

A very interesting phenomenon is the observation of non-trivial
correlations in the $A_2^{-+}$ sector. Asymptotically in $T$ this
mass should be given by
\begin{equation}
m_{A_2^{-+}}\;\to\; 2\sqrt{\mel^2+\sin^2\left({2\pi\over N}\right)},
\label{eq:a2mass}\end{equation}
where $2\pi/N$ is the lowest non-zero mode for an electric gluon. This
goes to \mdeb{} only in the limit of the spatial volume going to
infinity ($N\to\infty$). This formula works surprisingly well even at
the temperature we study. For our lattice sizes ($N=16$), taking \mel{}
to be half the lowest $A_1^{++}$ mass, we find $m_{A_2^{-+}}\approx0.90$
for $\t=6$ and $1.06$ for $\t=4$.

It is also interesting that below
\tc, the $A_1^{++}$ mass originating from the $E^{++}$ representation of
$O_h$ seems to be degenerate with the $B_1^{++}$ mass originating from
the same representation. This is consistent with other evidence that the
zero temperature spectrum is realised dynamically in this phase. Further
investigation of these other glueball channels, at a sequence of
temperatures, would be rewarding.

\section{Conclusion}\label{sec:conclusion}

In this paper we have presented first measurements of screening masses
for pure-glue operators at high temperatures in $SU(3)$ gauge theories.
We have identified the symmetry of the `$z$-slice group', \sym, and
presented the reduction of loop operators under this symmetry. In order
to understand the dynamics giving rise to the measured screening masses,
we have also performed the reduction of these representations under the
symmetries of the zero-temperature theory. Furthermore, we have examined
the symmetries of the link operators, `gluons', and performed the
reduction of direct products of these representations.

Correlations in the $A_1^{++}$ channel were the least noisy and, hence,
the easiest to study. We were able to reach a detailed understanding of
the dynamics in this channel. Our observations are completely consistent
with the usual picture of a deconfinement phase transition in QCD. In the
low-temperature phase one finds a (strong) perturbation of the zero
temperature spectrum, whereas the high-temperature phase can be
understood in terms of multi-gluon exchanges. Of course, as is
well-known, at temperatures of $1.5\tc$, where our simulations were
performed, the deconfined theory is not weakly interacting.

In the confined phase the correlation functions were seen to dynamically
reproduce the symmetries of the zero temperature theory. We observed
this both in the spectrum and the `wavefunctions'--- the lowest mass in
the $A_1^{++}$ channel (of \sym) originated from the correlations of
operators transforming under the $A_1^{++}$ channel of the symmetry
group of the zero temperature theory, and the next excited state in this
channel could be identified as the correlations of the operators
transforming as the $E^{++}$ representation. As a result it is possible
to talk of thermal shifts in glueball masses for $T<\tc$. We found that
at $T=0.75\tc$ the mass of the $0^{++}$ glueball dropped to about half
its zero temperature value. This is a larger mass shift than anything
seen in the meson spectrum.

The dynamical behaviour of these correlations in the deconfined phase
are completely different. Both the mass-spectrum and the `wavefunctions'
indicate a splitting between correlations carried by the electric and
magnetic polarisations. Group theoretically, the $A_1^{++}$
representation of \sym{} can arise from either two electric or two
magnetic gluon exchanges in a relative s-wave state. Thus, to
${\cal O}(g^2(T))$, the screening mass in this channel gives either
$2\mel$ or $2\mag$. Such a splitting is seen in the spectrum, and is
correlated with the use of operators with and without time-like links.
Furthermore, the $A_1^{++}$ screening mass in the electric sector agrees
very well with the measured Wilson-line screening mass. Hence, we can write
\begin{equation}
m_{A_1^{++}}\;=\;2.8(4)T\;\approx\;2\mel\quad{\rm and}\quad
  m'_{A_1^{++}}\;=\;5.8(4)T\;\approx\;2\mag\qquad\qquad(T=1.5\tc).
\label{eq:result}\end{equation}
These statements come with the usual caveat. The non-perturbative
measurements of the two different kinds of screening masses in the
$A_1^{++}$ channel are valid at all temperatures, but the identification
of these with twice the gluon electric and magnetic masses is valid only
at high temperatures.

Within the context of perturbation theory, corrections to the
identifications made above arise from multi-gluon exchanges. Since direct
products of three gluons do not contain $A_1^{++}$, the ${\cal O}(g^3(T))$
corrections can be obtained by resummation. Of course, these and higher
corrections are not negligible at the couplings where our simulations are
performed. Such corrections result in the mixing of electric and magnetic
channels, as is seen easily when the perturbative diagrams are written out.
Similar arguments may explain the fact, a little surprising at first sight,
that \mag{} turns out to be larger than \mel. Their ratio is expected to be
proportional to $g(T)$, but we work at rather large values of $g(T)$; it is
entirely possible that at higher temperatures, and hence weaker couplings,
one obtains a different ordering of these masses.

The $A_2^{-+}$ channel is correlated through similar two-gluon exchanges
in a relative p-wave state. Thus, in perturbation theory at order
greater than $g^2(T)$ the screening mass is split from the $A_1^{++}$
channel. On lattices of finite spatial volumes this screening mass
is split from the $A_1^{++}$ even in the limit $g(T)\to0$. Surprisingly,
even for our simulations at $1.5\tc$, the observed splitting between the
screening masses in these two channels seems to be due entirely to this
finite volume effect. It would be extremely interesting to continue these
measurements to higher temperatures.

A cross check on a perturbative interpretation would be provided by an
unambiguous measurement of the $B_1^{++}$ and $B_2^{++}$ screening
masses. These are also correlated by the exchange of two magnetic
gluons, and hence should be (asymptotically) degenerate with the larger
mass measured in the $A_1^{++}$ channel. Present measurements do not
support this strongly, but measurements at larger physical distances and
at higher temperatures are clearly needed. We found the negative C-parity
correlations extremely noisy. Better techniques for signal enhancement
in these channels are clearly desirable, since the leading contributions
to these are due to three gluon exchanges.

In summary, we have performed measurements of screening masses in the
pure gauge sector of QCD at finite temperatures, and identified the main
change in dynamics across the phase transition temperature. In the low
temperature phase we have measured the thermal shift in the $0^{++}$
glueball mass. In the deconfined phase we have been
able to identify screening masses which, in the high temperature limit,
measure electric and magnetic gluon masses, and seen that they are not
degenerate. In a sense, these define `non-perturbative' gluon masses.
Further studies of these screening masses are clearly of major interest.

The work of UMH was supported in part by the DOE under grants
\#~DE-FG05-85ER250000 and \#~DE-FG05-92ER40742.  The research of F.K. was
supported in part by the National Science Foundation under Grant No.
PHY89-04035 and the German Research Foundation (DFG) under Grant Pe
340/3-2. UMH and FK would like to thank the Institute for Theoretical
Physics at UCSB for the kind hospitality  extended to them while the paper
was in the finishing stages. We would like to thank Uwe-Jens Wiese for
providing us with the updating routines used to generate the gauge field
configurations. The computations have been performed at the Supercomputer
Center HLRZ in J\"ulich and at the RWTH Aachen.

\end{document}